\documentclass[%
 reprint,
superscriptaddress,
nofootinbib,
 amsmath,amssymb,
 aps,
]{revtex4-1}
\usepackage{xcolor}
\usepackage{graphicx}
\usepackage{dcolumn}
\usepackage{bm}

\usepackage{rotating} 

\begin{document}

\preprint{APS/123-QED}

\title[Resonant neutrino self-interactions]{Resonant neutrino self-interactions and the $H_0$ tension}

   \author{Jorge Venzor}\email{jorge.venzor@cinvestav.mx}
    \affiliation{ Departamento de F\'isica, Centro de Investigaci\'on y de Estudios Avanzados del I.P.N. 
                    Apartado Postal 14-740, 07000, Ciudad de M\'exico, M\'exico}
     \affiliation{Tecnologico de Monterrey, Escuela de Ingenier\'ia y Ciencias,
Av. Heroico Colegio Militar 4700, 31300, Chihuahua, Chihuahua, M\'exico}
       \author{Gabriela Garcia-Arroyo}\email{arroyo@icf.unam.mx}
    \affiliation{Instituto de Ciencias F\'isicas, Universidad Nacional Aut\'onoma de M\'exico, Apartado Postal 48-3, 62251 Cuernavaca, Morelos, M\'exico}
    \affiliation{Departamento Ingenier\'ia Civil, Divisi\'on de Ingenier\'ia, Universidad de Guanajuato, Guanajuato, C.P. 36000, M\'exico.}
    \author{Josue De-Santiago}\email{Josue.desantiago@cinvestav.mx}
    \affiliation{ Departamento de F\'isica, Centro de Investigaci\'on y de Estudios Avanzados del I.P.N. 
                    Apartado Postal 14-740, 07000, Ciudad de M\'exico, M\'exico}
    \affiliation{Consejo Nacional de Humanidades, Ciencias y Tecnolog\'ias,
    			Av. Insurgentes Sur 1582, 03940, Ciudad de M\'exico, M\'exico}
 \author{Abdel P\'erez-Lorenzana}\email{aplorenz@fis.cinvestav.mx}
    \affiliation{ Departamento de F\'isica, Centro de Investigaci\'on y de Estudios Avanzados del I.P.N. 
                    Apartado Postal 14-740, 07000, Ciudad de M\'exico, M\'exico}


\date{\today}

\begin{abstract}

In this work, we study the previously unexplored resonant region of neutrino self-interactions.
Current disagreement on late and early time observations of the Universe expansion could be solved with  
new physics acting before the recombination era. Nonstandard neutrino self-interactions are among the most appealing candidates to solve this issue since they could be testable in the near (or midterm) future.
We use linear cosmological datasets to test neutrino self-interactions for a sample of fixed scalar mediator masses in the range $10^{-2}$ eV $\leq m_{\varphi}\leq 10^{2}$ eV.
The resonant behavior produces observable effects at lower couplings than those reported in the literature for heavy and light mediators.
We observe that in the best case scenario, using the Planck + BAO dataset, the tension with local measurements of $H_0$ eases from 4.9$\sigma$ (for $\Lambda$CDM) down to 2.8$\sigma$.
Albeit, this is driven mainly by the addition of extra radiation, with $\Delta N_{\rm eff}\sim 0.5$.
The joint dataset which includes Planck, BAO, and $H_0$ prefers a nonzero interaction from 2.3$\sigma$ to 3.9$\sigma$ significance in the range $0.5$ eV $\leq m_{\varphi}\leq 10$ eV.
Although, this last result is obtained with data that are still in tension.
These results add the last piece in the parameter space of neutrino self-interactions at the linear perturbation regime.
\end{abstract}

\maketitle


\section{\label{sec:introduction} Introduction}

Decades of observational advances have led to an estimate of the expansion rate of the Universe, $H_0$, to within one percent of accuracy.
The local measurements with supernovae calibrated with Cepheids lead to a value of $H_0=73.04\pm 1.04$ km s$^{-1}$ Mpc$^{-1}$ \cite{Riess_2022,Riess:2022mme},
while early Universe measurements of the cosmic microwave background (CMB) radiation and the baryonic acoustic oscillations (BAO) consistently lead to a smaller value \cite{Aghanim2018Planck,ACT020,DES2018}, leading to a discrepancy of around $5\sigma$ (see for instance  \cite{Verde2019,Riess_2021,Riess2019ApJ,H0licow2020,Riess_2022,Riess:2022mme,Krishnan_2020}).
The origin of this tension is still unknown and observational systematic errors have not been completely ruled out, although, there is no clear evidence that systematics could erase the tension \cite{Rameez_2021,philcox2022determining,Mortsell_2022,Philcox2022}.
Thus, if this tension stands it will require new physics to explain it.

While there are models that have tried to solve the tension at late epochs (see for example \cite{Alestas2021,Alestas2022,Heisenberg2022,Perivolaropoulos2021,Giampaolo2020,Camarena2021,Efstathiou2021}), CMB measurements of $H_0$ certainly depend on the cosmological model, thus, it seems more natural to rather introduce new physics beyond the base $\Lambda$CDM model at early epochs.
Early Universe models may include modifications on the dark components, such as early dark energy (EDE) \cite{Niedermann2021,Boylan_Kolchin_2021,Reeves_2023}, or dark radiation \cite{Brinckmann2023}, among others \cite{Di_Valentino_2021,SCHONEBERG20221}.
Here, one of those possible models is the one considering neutrino nonstandard self-interactions (NSIs).
The main advantage of neutrinos, compared to other models, is that the framework contains a small number of unknown parameters that are being, or will be, explored in terrestrial, solar, astrophysical, and other cosmological environments and there is no need for fine-tuning.
That is, it is easier to test neutrino NSI models compared with the same task for dark sector ones \cite{Heurtier2017,Huang2018,Brune2019,Das2021,Schoneberg2019,Brdar2020,Venzor2021,Esteban2021_probing,Cerdeno2021,Deppisch2020,Bustamante2020,Akita2022,Dutta:2023fdt,Medhi:2021wxj,Ge:2021lur,Escrihuela:2021mud,Escudero2020_CMB_search,Lyu2021,Seto2021,Fiorillo2022,chang2022towards}.

The dynamics of the neutrino NSIs heavily depends on the nature of the mediator particle, particularly on its mass.
On the one hand, an interaction mediated by a very light particle (compared with the temperature of the Universe) becomes more important at late times, the information about the mediator mass is erased by the interaction's nature, and strong bounds are obtained for the interaction coupling \cite{Forastieri2015,Forastieri2019,Venzor2022}.
On the other hand, a heavy mediator implies a tight coupling at early times that fades away faster than the expansion of the Universe.
In the latter case, observables depend on the effective coupling $(g_{\nu}/m_{\varphi})^2$ which in some works a bimodal posterior distribution has been observed . One of its peaks prefers a large positive coupling, that deviates from the standard neutrino behavior \cite{Kreisch2020,Park2019,kreisch2022atacama,das2023magnificent}. 
Although, other independent analyses have not observed this peak with a good statistical significance \cite{brinckmann2020self,Choudhury2021}.
Because the required effective coupling is large, compared to the Fermi constant, the cosmological and astrophysical relics may constrain this scenario \cite{Bell2006,Archidiacono2014,CyrRacine2014,Oldengott_2015,Oldengott_2017,Lancaster2017,Kreisch2020,Park2019,Choudhury2021,Blinov2019,brinckmann2020self,mazumdar2022flavour,Huang2021,kreisch2022atacama,das2023magnificent,RoyChoudhury2022}.

So far, the cosmological literature lacks the study of intermediate-mass range mediator particles. 
In this case, neutrino scattering peaks at the epoch when acoustic oscillations were shaped, between redshifts $10^6$ and $10^3$, or at neutrino temperatures of about 100 eV well down to 0.1 eV.
This translates to a resonant region of interest corresponding to a mediator mass that can be located within that same range.
Because of the resonance, the interaction rate gets enhanced and similar phenomenological implications as those observed in the light and heavy mediator cases could be reached with a much smaller coupling $g_{\nu}$.
Resonant neutrino self-interaction implications have been previously studied in other astrophysical environments \cite{Creque2021,chang2022towards,cerdeno2023constraints}. 
In this paper, we study the previously unexplored cosmological implications of the resonant region.

For our purpose, we shall model neutrino self-interactions with a Breit-Wigner cross section and compute its thermal average over all available energies at any given neutrino temperature.
Then, we will compute the interaction rate as a numerical function of the redshift for a fixed mediator mass.
Additionally, we shall employ a modified version of the linear perturbation solver \textsc{CLASS} \cite{class2,class4} and \textsc{MontePython} \cite{Montepython2013,Brinckmann2019_montepython} among the latest public available cosmological datasets to constrain the parameters of the model for a sample of fixed mediator masses in the range $10^{-2}$ eV $\leq m_{\varphi}\leq 10^{2}$ eV.
We then discuss the implications on the cosmological parameters of the resonant neutrino self-interactions.
Finally, we compare our results with the latest parameter constraints and use Bayesian model comparison tests.

The rest of this manuscript is organized as follows.
Along Sec. \ref{sec:scatt_rate}, we numerically compute the resonant interaction rate of the neutrino self-interactions.
In Sec. \ref{sec:class} we present the perturbation spectra for linear cosmology for different mediator masses.
Model parameter constraints and their implications are discussed within Sec. \ref{sec:constraint}.
Our conclusions are highlighted along Sec. \ref{sec:conclusions}. 
Finally, an appendix providing further details regarding cosmological constraints to the model has been added to the end.

\section{\label{sec:scatt_rate} Neutrino resonant scattering rate}

We start our discussion by addressing the computation of the scattering rate for a neutrino self-interacting in the early Universe.
Neutrino nonstandard self-interactions (NSIs) mediated by a hypothetical particle affect their free-streaming.
In this paper, we only consider $\nu \nu \rightarrow \nu \nu $ elastic scattering and its effects, assuming that the interaction is mediated by a scalar particle.
Yet, we should stress that the analysis could be extended to interactions mediated by higher spin particles with a proper rescaling.

We model the interaction with a Breit-Wigner cross section
\begin{equation}\label{eq:breit}
    \sigma (s)=\frac{g_{\nu}^4}{4\pi}\frac{s}{[s-m_{\varphi}^2]^2+\Gamma_{\varphi}^2m_{\varphi}^2},
\end{equation}
in which we are assuming massless neutrinos. 
This approach is model-independent.
In above, $g_\nu$ stands for the self-interacting coupling, $s=E_{\rm CM}^2=4E_{\nu}^2$ is the center of mass Mandelstam variable, $m_{\varphi}$ is the scalar mediator mass, and $\Gamma_{\varphi}=g_{\nu}^2m_{\varphi}/4\pi$ corresponds to the decay width. 
This cross section peaks at $E_{\nu}\approx m_{\varphi}/2$, which would be later translated to a peak at $T_{\nu}\sim m_{\varphi}$.
Notice that above Eq. (\ref{eq:breit}) reduces to the high and low energy expected limits, $\sigma(s\gg m_{\varphi}^2)\approx g_{\nu}^4/(4\pi s)$ and $\sigma(s\ll m_{\varphi}^2)\approx (g_{\nu}^4 s)/(4\pi m_{\varphi}^4)$.
In the zero neutrino width limit the cross section reduces to a Dirac delta function
\begin{equation}\label{eq:dirac}
    \lim_{\Gamma_{\varphi}\rightarrow 0} \sigma(s)=\frac{g_{\nu}^2 \pi}{m_{\varphi}^2}s \delta (s-m_{\varphi}^2).
\end{equation}
This is a very good approximation for the relevant range of parameters where $m_{\varphi}\leq 100$ eV and $g_{\nu}<10^{-10}$.
Also, notice that around the resonance the cross section becomes proportional to $g_{\nu}^2$, thus getting substantially enhanced.

In cosmology, since the neutrino bath energy is not monochromatic, all available energies need to be taken into account through the thermally averaged cross section.
Thus, in order to compute the interaction rate of the $\nu \nu \rightarrow \nu \nu$ scattering, we use
\begin{equation}\label{eq:def_rate}
    \Gamma_{\rm scatt}=\left< \sigma v_{\rm MOL}\right> n_{\nu},
\end{equation}
where $v_{\rm MOL}$ is the M\o ller velocity and $n_{\nu}$ stands for the neutrino number density.
As the resonant peak occurs much later than the standard neutrino decoupling, it is safe to assume that the neutrino number density evolves as in the standard case, thus, $n_{\nu}=3\zeta(3)T_{\nu}^3/(2\pi^2)$.
Here, we follow the framework developed in refs. \cite{GONDOLO1991,Vassh2015}.
Therefore,  the thermally averaged cross section can be written as
\begin{equation}\label{eq:moller}
    \left< \sigma v_{\rm MOL}\right>= \frac{1}{n_\nu^2}\int \frac{d^3p_1}{(2\pi)^3}\int \frac{d^3p_2}{(2\pi)^3} f(p_1) f(p_2) \sigma(s) v_{\rm MOL},
\end{equation}
where $f(p)=(e^{p/T}+1)^{-1}$ is the equilibrium neutrino distribution, which we assume is preserved even when the resonant NSI is near its peak. 
After some redefinitions, Eq. (\ref{eq:moller}) can be reduced to
\begin{equation}
    \left< \sigma v_{\rm MOL}\right>=\frac{T^2}{16 \pi^4n_\nu^2}\int_0^{\infty} s\sigma(s)F(s;T)ds,
\end{equation}
where $F$ is a numerical function defined as
\begin{equation}\label{eq:vassh}
    F(s;T)= \int_{\sqrt{s}/T}^\infty\frac{dx\ e^{-x}}{1-e^{-x}}G(x;T).
\end{equation}
Here, the function $G$ is respectively given as
\begin{equation}
    G(x;T)=A+\ln \left( \frac{1+e^{-(x/2+A)}}{1+e^{-(x/2-A)}} \right),
\end{equation}
where $A=\sqrt{x^2- s/T^2}/2$.
Plugging Eqs. \eqref{eq:dirac} and \eqref{eq:vassh} into Eq. \eqref{eq:def_rate} results in the final expression
\begin{equation}\label{eq:scatt_rate}
    \Gamma_{\rm scatt}= \frac{g_{\nu}^2 m_{\varphi}^2}{24 \pi \zeta(3) T_\nu}F(m_{\varphi}^2;T_\nu).
\end{equation}

Furthermore, it is worth mentioning that we have compared the above scattering rate with the one obtained by using the nonapproximated expression \eqref{eq:breit}, where, to have a good numerical result, it is necessary to numerically sample it with very high precision around the resonance. 
We found that both expressions \eqref{eq:breit} and \eqref{eq:dirac} are indistinguishable 
as long as the temperature is close to the resonance peak.
In this work, we will use the numerical solution given by Eq. \eqref{eq:scatt_rate} in the cosmological linear perturbation regime analysis.

\begin{figure}
    \centering
    \includegraphics[width=.47\textwidth]{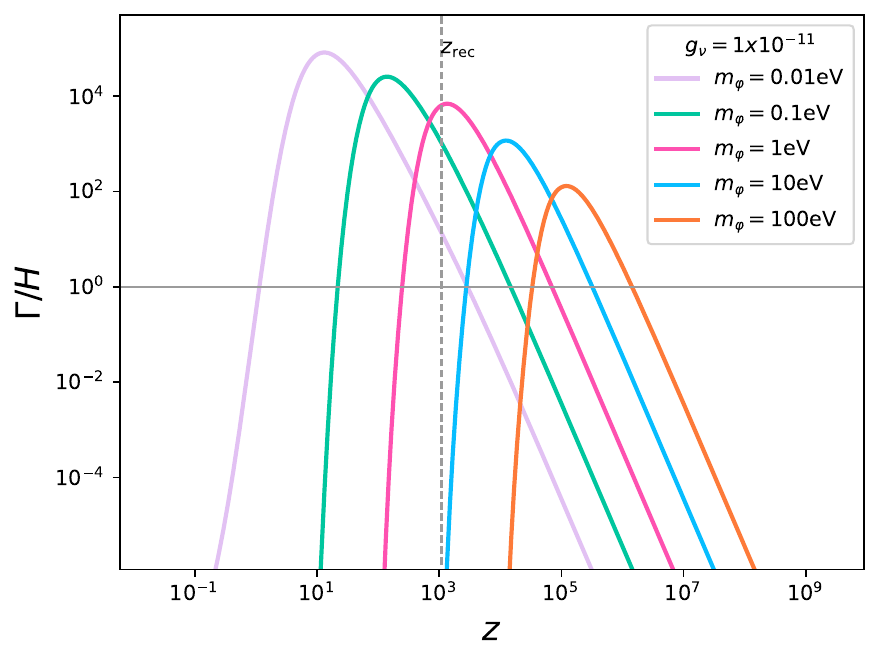}
    \caption{Interaction rate over the Hubble factor as a function of redshift. Colors denote different values of the mediator mass $m_{\varphi}$, for a fixed $g_{\nu}$ value. The horizontal line denotes the equality condition between $\Gamma$ and $H$, and the vertical dashed line shows the redshift at the recombination era. The baseline cosmological parameters were fixed according to Planck's 2018 results.}
    \label{fig:gamma_mphi}
\end{figure}

The resonant scattering rate is a smooth function that peaks around temperatures $\sim m_{\varphi}$ or at a redshift
\begin{equation}\label{eq:redshift_peak}
    z_{\rm peak}\sim 4 \times 10^3\left(\frac{m_{\varphi}}{{\rm eV}} \right) \, .
\end{equation}
For our analysis, we shall choose a sample of different fixed masses that lie between $10^{-2}$ eV and $10^2$ eV, or equivalently, consider scattering rate peaks that span from the redshift $10^{5}$ to $10$.
Fig. \ref{fig:gamma_mphi} shows that, indeed, we have a smooth interaction rate with a maximum around the resonance, which only depends on $m_\varphi$. Varying $g_\nu$ has the overall effect of globally scaling the interaction.
Also, the peak on the resonance appears at the redshift dictated by Eq. \eqref{eq:redshift_peak}, as a consequence, the temporal behavior of the interaction would only depend on the mediator mass.

As the physics of neutrino self-interactions has stronger effects in cosmology if it occurs before recombination,
we can check when the interaction rate surpasses the Hubble function for a fixed coupling value.
In Fig. \ref{fig:gamma_mphi} we plot the ratio $\Gamma/H$ for a relatively small coupling.
There, we can observe that lighter mediator's self-interactions enter into equilibrium too late, while heavier mediator's ratio is suppressed by the mass value.
In contrast, intermediate masses promote the entrance into equilibrium for the interaction well above the recombination era.
Notice that for a mediator mass of about $1$ eV, the peak of the rate-to-Hubble ratio lays just above recombination redshift $z_{rec}$.
Thus, we could expect to have stronger effects on the cosmological observables with medium-range masses.
Empirically, we found that the height of the peak is similar for different masses when the relation $g_\nu/m_{\varphi}^{0.4}\sim
 c$, where $c$ is a constant, is satisfied.

Finally, as we mentioned earlier, the narrow Breit-Wigner is not a very good approximation to describe the low and high energy tails of the interaction rate.
However, as the tails are negligible compared to the resonance since they go as $g_\nu^4$, thus, they are not expected to be relevant for any region of the parameter space.
Nonetheless, we must mention that we have tested the resonant interaction plus the low and high energy tails versus the resonance alone and found no differences between both.

During this project, we neglect the contribution of the scalar field fluid in the cosmological evolution, however, this is a good approximation due to the smallness of $g_\nu$-values that we will be working with hereafter.

\subsection{\label{sec:perturbations} Self-interaction in cosmological perturbations}

In this subsection, we will discuss the role of the resonant interaction rate in cosmological perturbations.
If $\Psi$ describes the perturbations to the background distribution function $f(x,q,t)=f_0(q(t))(1+\Psi(x,q,t))$,
the neutrino evolution depends on this perturbation function.
At early epochs due to the smallness of the NSIs, the neutrinos decoupled from the cosmic plasma as in the standard case around $T_{\nu}\sim 1$ MeV.
Afterward, neutrinos free-stream until the resonant self-interaction becomes relevant.
We employ the formalism of the Boltzmann hierarchy equations~\cite{Ma1995} in conformal Newtonian gauge  (with $\Gamma=\Gamma_{\rm  scatt}$) and the relaxation time approximation. These considerations result in the system of equations, for the multipole Legendre moments of $\Psi$ \cite{Hannestad2000,Archidiacono2014,Forastieri2015}, 
\begin{eqnarray} 
\dot{\Psi}_0 &=& -\frac{qk}{\epsilon}\Psi_1- \dot{\phi}\frac{d \ln f_0}{d \ln q} \, , \\
\dot{\Psi}_1 &=& \frac{qk}{3\epsilon}(\Psi_0 -2\Psi_2) -\frac{\epsilon k }{3q}\psi \frac{d \ln f_0}{d \ln q}\, ,\\
\dot{\Psi}_2 &=&\frac{qk}{5\epsilon}(2\Psi_1-3\Psi_3) -a\Gamma \Psi_2\, , \label{eq:boltzmann_l2}\\
\dot{\Psi}_{l\geq 3} &=& \frac{q k}{(2l+1)\epsilon}\left(l\Psi_{l-1}-(l+1)\Psi_{l+1}\right) -a \Gamma \Psi_l\, , 
\end{eqnarray}
where $\phi$ and  $\psi$ stand for the corresponding gravitational scalar potentials, $f_0$ is the unperturbed Fermi-Dirac distribution, $\epsilon = \sqrt{q^2 +a^2m_{\nu}^2}$ is the comoving energy and $q$ is the magnitude of the comoving momentum. 
Notice that $l=0$ and $l=1$ are free from the collision term due to energy and momentum conservation of the elastic $\nu \nu \rightarrow \nu \nu$ process.

Although, while computing the interaction rate we have ignored the neutrino mass, given its relevance to cosmological neutrino dynamics, we are going to consider massive neutrinos when analyzing the data.
Accordingly, we shall be using the cosmological parameter $\sum m_{\nu}$ hereafter.

The system of Boltzmann linear equations for all the matter components must be simultaneously solved, jointly to the perturbed  Einstein equations, whose right--hand side is concomitantly determined by the perturbed matter variables, which are calculated from the integration of the multipole moments $\Psi_{\ell}$.
In particular, the anisotropic stress is defined through the integration of the second moment, $\ell =2$, over momenta.
As the modification of the Boltzmann hierarchy takes place just for $\ell \geq 2$,  Eq. \eqref{eq:boltzmann_l2}  creates a change in the  neutrinos anisotropic stress, which in turn modifies the  Einstein shear equation:
\begin{equation}\label{eq:shear}
    k^2(\phi - \psi)= 12 \pi G a^2(\rho + P)\Sigma\, .
\end{equation}
Here $\rho$, $P$, and $\Sigma$ are the total density, pressure, and anisotropic stress coming from all the different matter sources.
This equation couples together all the matter components as their evolution sources the scalar potentials $\phi$ and $\psi$ which, in turn, influence the evolution of each matter component.
In the case of neutrinos, their free-streaming originates anisotropic stresses $\Sigma_{\nu}$,
but the NSIs can delay or suppress the said free-stream.
This modification alters Eq. \eqref{eq:shear}, that in turn, changes the evolution of all the matter perturbations,
including CMB photons and baryons.
These effects are the reason why cosmology data becomes sensitive to the neutrino NSI.

\section{\label{sec:class} Perturbation spectra and parameter exploration }

\begin{figure}
    \centering
    \includegraphics[width=.48\textwidth]{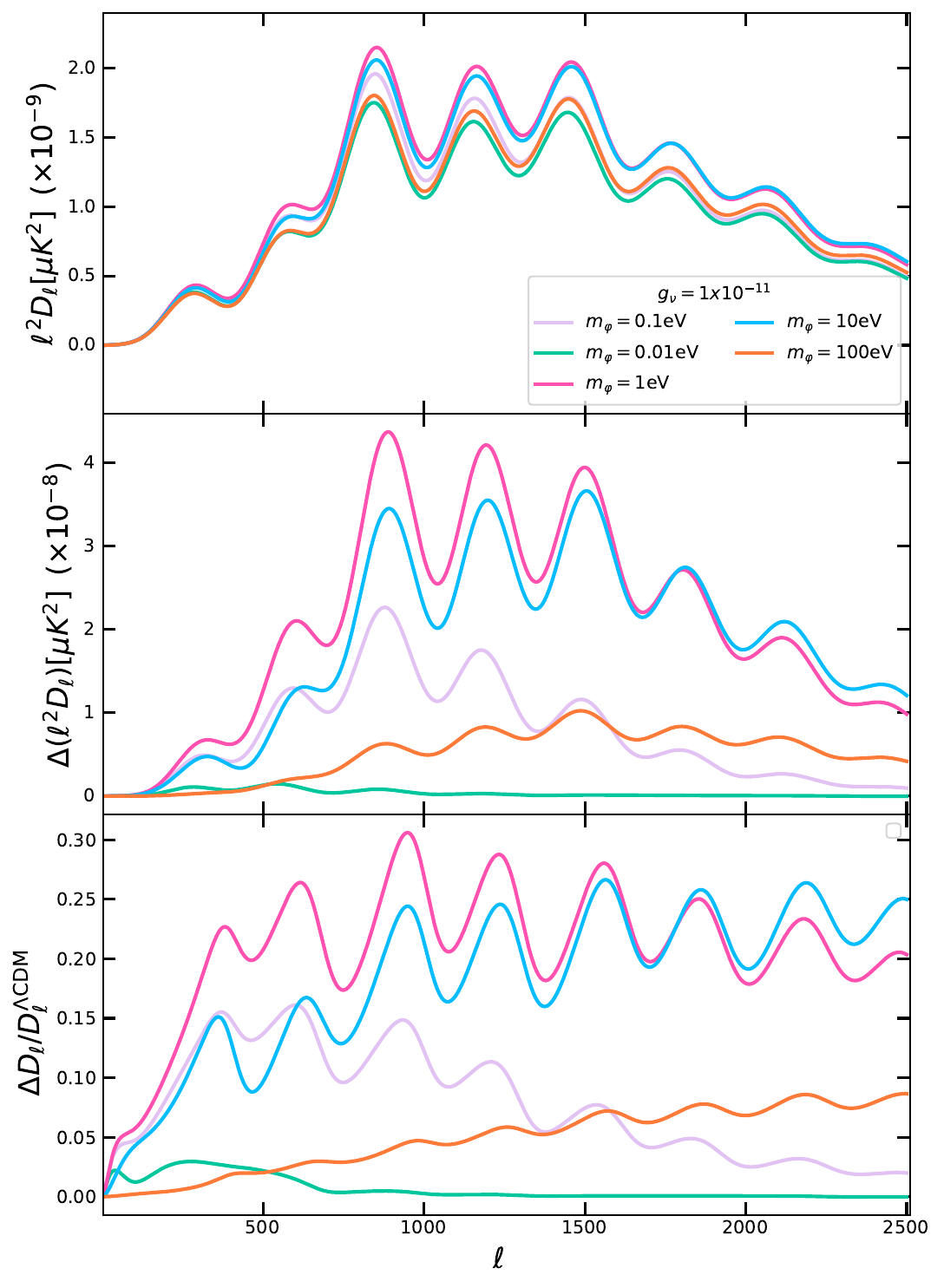}
    \caption{Effects of the particle mass mediator $m_{\varphi}$ on the CMB TT power spectra for a fixed $g_{\nu}$ value. The colors correspond to five different masses as specified by the labels. In the upper panel the spectra are multiplied by $\ell^{2}$ just for visualization purposes, in the middle panel $\Delta D_{\ell}=D_{\ell}^{\rm res}-D_{\ell}^{\Lambda \rm CDM}$ is for the difference with respect $\Lambda {\rm CDM}$, since the bottom panel shows the residual plot. }
    \label{fig:TT_resonance_mphi}
\end{figure}

In this section, we discuss the cosmological spectra in the presence of a resonant NSI. 
CMB power spectra are sensitive to the resonant NSI, specifically, neutrino free-streaming suppresses photon fluctuations and shifts their phase.
Therefore, when NSIs reduce neutrino free-streaming, they naively enhance the spectrum.
This interaction is redshift dependent and carries two extra parameters $g_{\nu}$ and $m_{\varphi}$ that regulate the free-streaming scale and its local effects. Here we vary both parameters while fixing all others, in particular, baseline $\Lambda$CDM parameters are fixed to Planck 2018 results \cite{Aghanim2018Planck},  and $\Sigma m_{\nu}=0.06 {\rm eV}$ and $N_{\rm eff}=3.046$.
To compute the perturbation spectra we employ a modified version of the code \textsc{CLASS} \cite{class2, class4}.

We first explore the order of magnitude in which the coupling shows observable and reasonable effects.
As we mentioned in the previous section, the resonance naturally enhances the interaction in a window estimated by Eq. \eqref{eq:redshift_peak}.
The first interesting feature of the resonant NSIs is that we can observe a strong enhancement of matter and temperature spectra for much lower $g_{\nu}$-values than the ones revisited on the heavy and light mediator approximations.

To address the effects induced by the mediator mass, in Fig. \ref{fig:TT_resonance_mphi} we plot, for a fixed $g_{\nu}$ value, the spectra in two mass regions; below and above 1 eV. There, we see that the highest amplification of the CMB TT spectrum corresponds to the mass of  $1\, \rm{eV}$. This agrees with the previously observed fact that for such a mass the peak of the interaction rate to Hubble ratio happens just right above recombination. This reduces the free-streaming just in the relevant epoch to the CMB spectrum, which prevents photon fluctuations from being suppressed. 

Fig. \ref{fig:TT_resonance_mphi} exhibits the behavior of the CMB fluctuations for $g_{\nu}=10^{-11}$, this value generates a visible and moderate contribution of the interaction.
For other $g_{\nu}$-values the global effects presented in Fig. \ref{fig:TT_resonance_mphi} will be the same and only will be enhanced or diminished.
From Fig. \ref{fig:gamma_mphi} we see that
masses above $1\, \rm eV$ mainly affect the high $\ell$ multipoles of the CMB, because the window where the interaction rate grows with respect to the Hubble factor takes place before recombination.
Also for the same reason, in this mass regime, the spectra have a less noticeable impact at low multipoles. On the contrary, the smaller masses will contribute more at low $\ell$ multipoles because its interaction rate is negligible compared to Hubble at early times and, as a consequence, we obtain deviations that decay at high multipoles.

 \begin{figure}
    \centering
    \includegraphics[width=.48\textwidth]{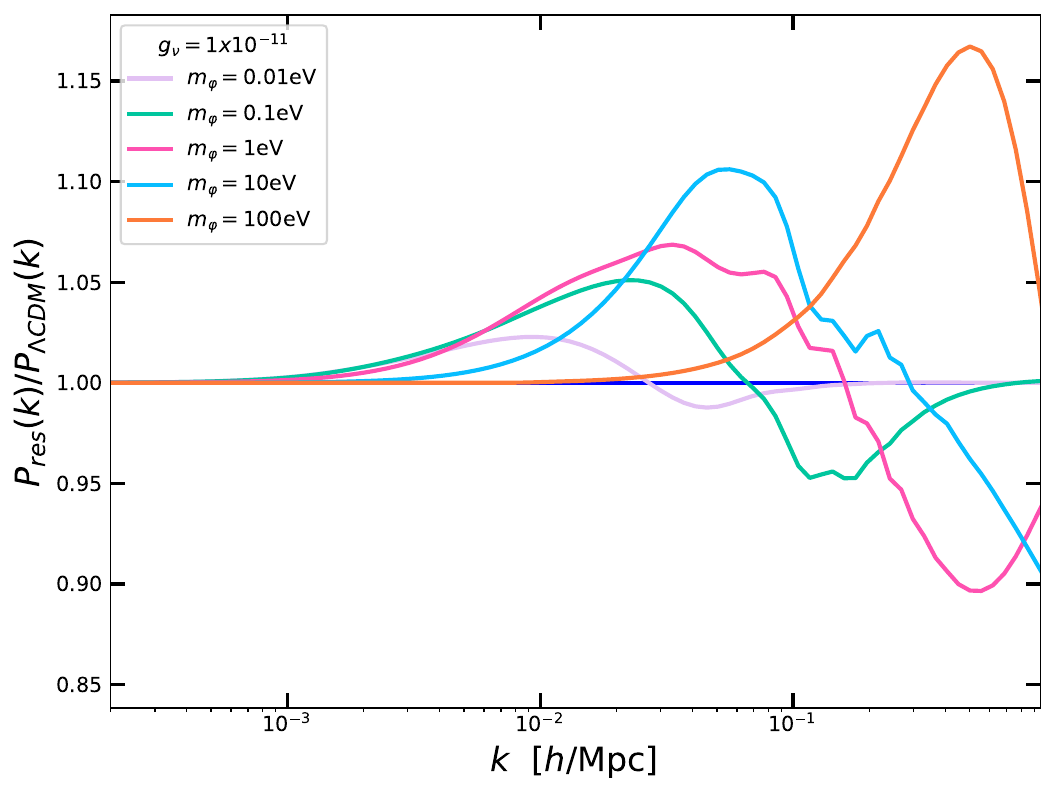}
    \caption{The ratio of the resonant NSI Matter Power Spectra compared to $\Lambda$CDM, evaluated at redshift $z=0$. The $g_{\nu}$ value is the same for the different masses of the mediator particle.}
    \label{fig:MPS_resonance}
\end{figure}

Dark matter perturbations are also modified by the NSIs. For instance, the interaction prevents free streaming and therefore reduces the neutrino contribution to the anisotropic stress. This, according to Eq. (\ref{eq:shear}), modifies the metric potentials, $\phi$, $\psi$, which observationally can be translated to modifications in the matter power spectrum (MPS). In Fig. \ref{fig:MPS_resonance} we present the MPS for masses in different regimens.
For all of them, we find that the large scales (small $k$) behave like $\Lambda$CDM. This happens because these modes enter the horizon after the resonance peak. The exact $k$ at which this happens do depend on the specific mediator mass. Bigger masses peak earlier, allowing the modes to mimic $\Lambda$CDM at larger $k$'s.

The behavior of the larger $k$ modes depends on the mass of the mediator;  for the heaviest explored masses we found that the carried effects on the MPS at those scales are similar to the results of \cite{Kreisch2020}. 
Thus, we focus here on the  MPS produced by masses below or equal to $1 \rm{eV}$, for which the global behavior is the same. The largest $k$ modes enter the horizon  before the interaction becomes relevant so we don’t have differences with respect to $\Lambda$CDM.
Recalling that, as the k-mode increases in size, it enters the horizon earlier. Consequently, intermediate scales are particularly susceptible to the interaction rate since their  horizon entry may coincide with a time when the interaction is relevant. In fact, as we go to smaller $k$ values  we have a small suppression of power (as a consequence of a smaller metric potential $\psi$ compared to $\Lambda$CDM) that once we move to smaller $k$ grows until it exceeds the MPS of $\Lambda$CDM and reaches a maximum (associated now with a boost of $\psi$). 
Finally, for the smaller $k$'s, the MPS behaves again like in $\Lambda$CDM.
In the next section, we are going to discuss the constraints of these effects with cosmological data.

\section{\label{sec:constraint} Cosmological results}

In this section, we use different sets of cosmological observations to constrain the parameters of the resonant neutrino self-interactions model.
The observations include CMB power spectra from Planck 2018, which consist of  the combined TT, TE, EE, low E, and lensing likelihoods
\cite{Planck:2018nkj,planck_likelihood}.
Baryonic acoustic oscillations (BAO) measurements from BOSS DR12 galaxies \cite{BOSS:2016wmc} and low-z BAO data from 6DF \cite{Beutler_2011} and MGS \cite{Ross_2015}.
Additionally, we also incorporate the most recent local measurement of $H_0$ by the SH0ES collaboration \footnote{Note that models that incorporate late-time solutions of $H_0$ are inconsistent with the approach of incorporating a prior for $H_0$ into the analysis \cite{Giampaolo2020,Camarena2021,Efstathiou2021}. However, that is no the case for our model. } \cite{Riess_2022}.
Throughout this paper, we only use two data combinations that we call Planck + BAO and Planck + BAO + $H_0$, respectively.

In order to test the resonant model with observations we run multiple
Markov chain Monte Carlo (MCMC) with the publicly available code \textsc{montepython} \cite{Montepython2013,Brinckmann2019_montepython}.
We vary all six $\Lambda$CDM base parameters.
Additionally, we include the two neutrino parameters $\sum m_\nu$ and $N_{\rm eff}$ plus the neutrino-scalar coupling $g_\nu$.
We are assuming a diagonal and universal coupling.
As we mentioned in section \ref{sec:scatt_rate}, we fix the mediator mass to find the numerical solution of the interaction rate,
which we supply precomputed to \textsc{class}.
For this reason, we cannot use $m_\varphi$ as a free parameter.
Our main sample consists of the following set of masses: \{$10^{-2}$, $10^{-1}$, 1, $10$, $100$\} eV. This sample covers the entire resonant parameter space.

Additionally, we ran MCMCs using the $\Lambda$CDM model with Planck + BAO dataset for reference of the standard case.
Finally, we used both linear and logarithmic priors for $g_\nu$, in both cases we found the same results that we discuss below.
Hereafter, we will only present the results when using the linear prior.

\begin{table*}
\tiny
  \centerline{
\begin{tabular} {| l | c | c c |c c| c c| c c| c c|}

\hline
Data: Planck + & BAO & BAO & BAO+$H_0$  & BAO & BAO+$H_0$ & BAO & BAO+$H_0$ & BAO & BAO+$H_0$ & BAO & BAO+$H_0$\\
\hline
    $m_{\varphi}[\rm{eV}]$ & --- & \multicolumn{2}{c|}{ $10^{-2}$} &\multicolumn{2}{c|}{ $10^{-1}$} &\multicolumn{2}{c|}{ 1} &\multicolumn{2}{c|}{ 10}  &\multicolumn{2}{c|}{ 100} \\
    \hline
$H_0  [\rm{km\ s^{-1}Mpc^{-1}}]   $ & $67.2\pm 1.2               $ & $68.1\pm 1.4        $  & $71.12\pm 0.85           $ & $68.1\pm 1.4    $ & $71.22\pm 0.85   $ & $67.9\pm 1.3                $ & $71.18\pm 0.81         $ & $67.7^{+1.3}_{-1.4}      $ & $71.09\pm 0.81     $ & $67.5\pm 1.3      $ & $71.02\pm 0.81       $\\
$N_{\rm eff}                   $ &$2.95\pm 0.17              $& $3.12\pm 0.23       $  & $3.58^{+0.15}_{-0.17}    $ & $3.12\pm 0.22   $ & $3.59\pm 0.16    $ & $3.05^{+0.19}_{-0.22}       $ & $3.56\pm 0.15          $ & $3.03^{+0.20}_{-0.23}    $ & $3.54\pm 0.15      $ & $3.02\pm 0.21     $ & $3.54\pm 0.14        $\\
$\sum m_{\nu}[\rm{eV}]          $ &$< 0.0587                  $& $< 0.110               $ & $< 0.119                $ & $< 0.114                  $ & $< 0.123                  $ & $< 0.112                   $ & $< 0.110                  $ & $< 0.120                  $ & $< 0.117                    $ & $< 0.110               $ & $< 0.0915                     $\\    
$g_{\nu}\times 10^{14}     $ & 0 & $< 861                 $ & $< 909                  $ & $< 91.7                   $ & $< 106                    $ & $< 15.0                    $ & $12.4^{+4.4}_{-3.5}       $ & $< 36.7                   $ & $27\pm 10          $ & $< 147                 $ & $< 207                        $\\
$\Delta$AIC &2.06 & 5.66 &-6.60  & 4.98 &-10.20 &5.56 &-9.38 &6.02 & -4.82& 8.22& -6.56 \\
$H_0$ tension & $3.66\sigma$ &$2.83\sigma$&&$2.83\sigma$&&$3.09\sigma$&&$3.21\sigma$&&$3.33\sigma$&\\
\hline
\end{tabular}}
\caption{\label{tab:limits} Observational limits for different models with varying $N_{\rm eff}$ and $\sum m_\nu$.The upper limits are expressed at $2\sigma$ while the rest of the limits are at $1\sigma$.
The first column displays the results of the control chains varying $N_{\rm eff}$ without the interaction.
We obtain a positive value of the interaction coupling for $m_\varphi$ corresponding to 1 and 10 eV. We also computed the tension with the value from Cepheids which is reduced from its $4.9\sigma$ value for $\Lambda$CDM.}
\end{table*}

\begin{figure}
    \centering
    \includegraphics[width=.48\textwidth]{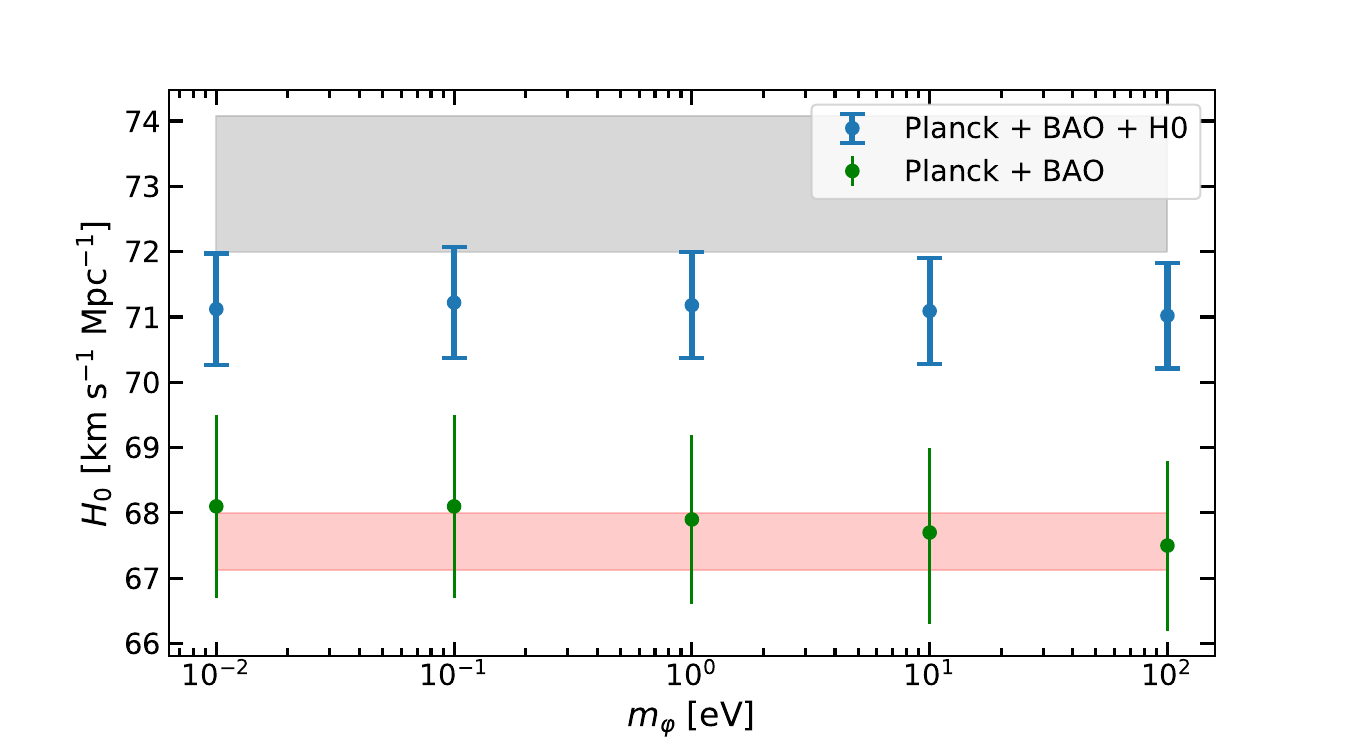}
    \caption{$H_0$ for the main sample of fixed mediator masses. Green points were obtained using the Planck + BAO  dataset, while the blue points also include the local measurement of $H_0$. The  gray band is the 1 sigma local measurement \cite{Riess_2022}, and red band corresponds our estimation using $\Lambda$CDM and Planck + BAO. }
    \label{fig:H0_vs_mass}
\end{figure}

In table \ref{tab:limits}, we show the parameter constraints obtained from the MCMC analysis for $H_0$ and the relevant neutrino parameters.
First, let us focus on the $H_0$ results,
as we can see in the table for the different masses and when using the Planck + BAO dataset the measurement of $H_0$ presents a larger uncertainty compared to the $\Lambda$CDM model ($H_0=67.56\pm 0.44 \ \rm{km\ s^{-1}Mpc^{-1}}$). 
This happens due to the wider parameter space. 
Furthermore, the best-fit estimation of $H_0$ is larger than that of $\Lambda$CDM for all masses, except for $m_\varphi$ = 100 eV. 
Taken together, these results lead to a reduction in the $H_0$ tension from 4.9$\sigma$ ($\Lambda$CDM) to $2.83\sigma$ ($m_{\varphi}=10^{-2}\rm{eV}$),  $2.83\sigma$ ($m_{\varphi}=10^{-1}\rm{eV}$),  $3.09\sigma$ ($m_{\varphi}=1\rm{eV}$),  $3.21\sigma$ ($m_{\varphi}=10^{1}\rm{eV}$), and $3.33\sigma$ ($m_{\varphi}=10^{2}\rm{eV}$).
Thus, the lightest masses reduce better the tension (see Fig. \ref{fig:H0_vs_mass}).
As reference cases, we also run chains varying $N_{\rm eff}$ without the interaction and another where $N_{\rm eff}=3.04$ is fixed while the interaction is switched on.
In the former case, the tension is reduced to $3.7\sigma$, which alone accounts for $\sim 57\%$ of the tension reduction.
While in the latter, the tension only reduces to $4.6\sigma$ (for $m_{\varphi}=1$ eV), which is consistent with our overall results.
These results show that the RNSIs do indeed help to reduce the Hubble tension, however, the model requires extra radiation for this task, this is similar to what it was previously observed in the heavy mediator approximation \cite{Kreisch2020,Park2019,Choudhury2021,brinckmann2020self,kreisch2022atacama}.

When we add the $H_0$ measurements into the batch, the model takes high $H_0$ values (see Fig. \ref{fig:H0_vs_mass}).
We observe that the preference for larger $H_0$ correlates with a larger $N_{\rm eff}$ (see table \ref{tab:limits} and figures \ref{fig:triangular100ev}--\ref{fig:triangular10minus2ev} in Appendix A).
We introduced the extra radiation via the parameter $N_{\rm ur}$ which corresponds to noninteracting radiation and $N_{\rm eff}=3.046+N_{\rm ur}$.
As we can see in table \ref{tab:limits}, the amount of extra radiation compensating for the lack of free-streaming is roughly $N_{\rm ur}\sim 0.5$.
Studies in both the heavy a light mediator approximations \cite{Kreisch2020,Park2019,Choudhury2021,brinckmann2020self,kreisch2022atacama,Forastieri2019,Venzor2022}
also show a strong positive correlation between $H_0$ and $N_{\rm eff}$.

\begin{figure}
    \centering
    \includegraphics[width=.44\textwidth]{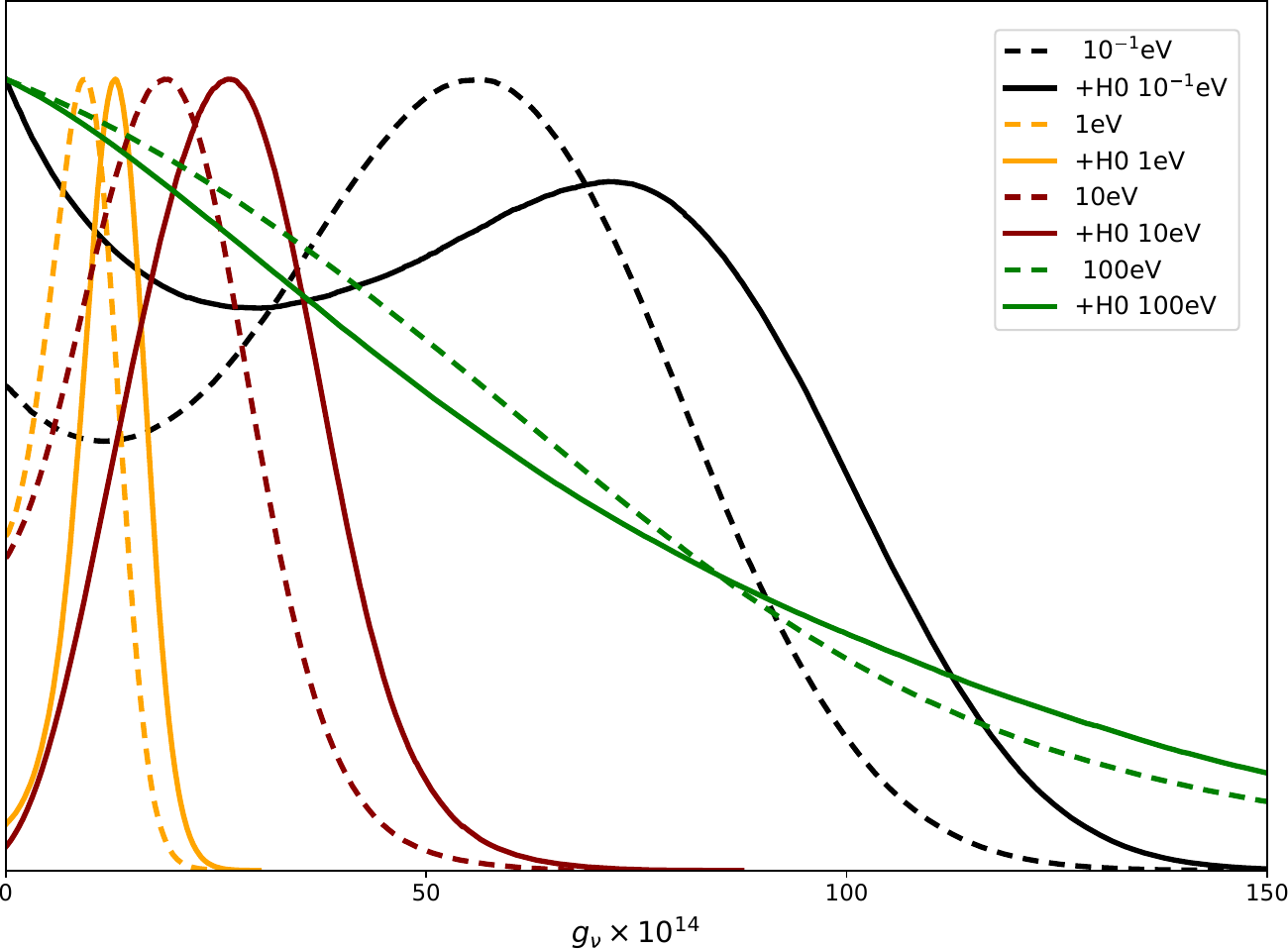}
    \caption{Posterior probability for the parameter $g_\nu$ at different masses of the field $\phi$. The dotted lines correspond to Planck+BAO constraints and the continuous lines to Planck + BAO + $H_0$.}
    \label{fig:gnu_linear}
\end{figure}

Now, let us discuss the results of the interaction coupling $g_\nu$.
In Fig. \ref{fig:gnu_linear} we show the posterior distribution for $g_\nu$ with both datasets.
The two datasets indicate that the coupling is more constrained for the intermediate masses of 1 and 10 eV, which is consistent with our analysis in Sec. \ref{sec:class} where we showed that these masses led to the strongest effects on the CMB.
However, the other masses also present strong upper limits on $g_{\nu}$. 
The weakest bounds were found for the heaviest and lightest masses. For instance, for $m_\varphi=10^{-2}$eV, 
 we found $g_\nu<8.61\times 10^{-12}$ at 95\% C.L. Still this bound is much stricter than the bounds found in the literature for light or heavy mediators, which are around $10^{-7}$ at best.

\begin{figure}
    \centering
    \includegraphics[width=.49\textwidth]{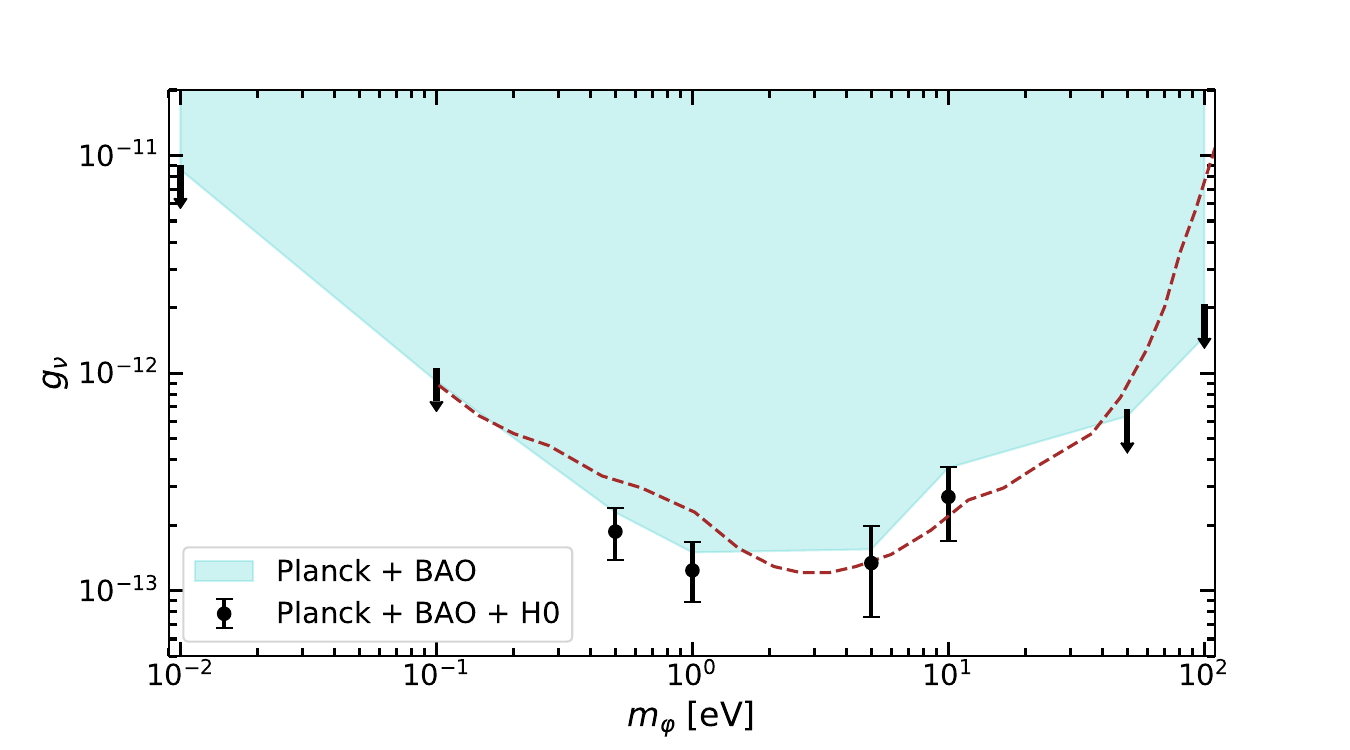}
    \caption{Coupling vs mediator mass parameter space. The cyan region reflects the excluded region for $g_\nu$ at 95\% C.L. using the Planck + BAO dataset. The black points represent either the upper bounds at 95\% C.L. (arrows pointing down) or the 1 sigma measurements of $g_\nu$ (error bars) for Planck + BAO + $H_0$. The dashed brown line indicates the bound from scalar decaying into neutrinos in a Majoron model \cite{Escudero2020_CMB_search}.  }
    \label{fig:gnu_vs_mass}
\end{figure}

When using the whole data pool the results on $g_\nu$ become more intriguing.
For the intermediate masses of 1 and 10 eV, the data prefer a nonzero interaction with a significance larger than $2\sigma$.
Considering this, we explored more masses in the neighborhood given by $m_\varphi=$\{$0.5$ eV, $5$ eV, $50$ eV\} (for a full discussion on these points see Appendix \ref{appendix} and table \ref{tab:limits2} therein).
In Fig. \ref{fig:gnu_vs_mass}, we show the estimations of $g_\nu$, the range where we have at least a 2.3$\sigma$ preference for nonzero values is $0.5$ eV $\leq m_{\varphi}\leq 10$ eV.
Specifically, the statistical preference for nonnull values are $3.90\sigma$ ($m_{\varphi}=0.5\rm{eV}$),  $3.54\sigma$ ($m_{\varphi}=1\rm{eV}$),  $2.31\sigma$ ($m_{\varphi}=5\rm{eV}$), and $2.70\sigma$ ($m_{\varphi}=10\rm{eV}$) respectively.
Here, it is important to state that this result is obtained with cosmological data that are still in tension, although the $\chi^2_{\rm min}$ is almost identical for RNSI (with $H_0$ included) and $\Lambda$CDM (Planck+BAO only).
These measurements expand around $7.6 \times10^{-14}\lesssim g_\nu \lesssim 37 \times 10^{-14}$, and would be difficult to test with experimental neutrino probes.
However, we argue that in astrophysical and cosmological setups the task becomes easier, as in recent years, many phenomenological ideas have led to new bounds.
For instance, a strong bound close to ours was found for a Majoron-like model\footnote{In these kind of models, the neutrino mass typically comes from a seesaw mechanism (see for example \cite{Chikashige1980,Gelmini1981}).} where the smallness of the coupling is naturally achieved \cite{Escudero2020_CMB_search}. 
New neutrino UV complete models should consider our results \cite{Berbig2020,He_2020,Lyu2021}.

Moreover, we discuss our constraints on the sum of neutrino masses. 
Overall the neutrino mass bounds are similar to the standard neutrino model. 
Except in one case, with $m_\varphi=100$ eV, using the complete dataset the neutrino mass bound becomes more restrictive $\sum m_\nu<0.09$ eV, which slightly overlaps with the minimum sum expected in the inverted hierarchy $\sum m_{\nu}^{\rm IH}>0.0986 \pm 0.00085$ eV \cite{Loureiro2019}
(For more details and constraints on other cosmological parameters, see appendix \ref{appendix}).

Finally, to evaluate the goodness of fit of the model, we employ the Akaike information criterion (AIC) \cite{AIC,AIC2}, which is defined as
\begin{equation}\label{eq:aic}
\rm{AIC}= -2\ln \mathcal{L}_{max}+2n_{free}\, ,
\end{equation}
where $\mathcal{L}_{\rm max}$ is the maximum likelihood for a model, and $\rm n_{free}$ the number of free parameters.
The AIC penalizes overfitting caused by extra free parameters. 
Thus, we define $\Delta \rm{AIC}= \rm{AIC}_{\rm{RNSI}}-\rm{AIC}_{\rm{\Lambda CDM}}$.
Here, a positive $\Delta \rm{AIC}$ favors the standard model and a negative value the RNSI model.
Furthermore, if the value of $\Delta$ lies in the interval $[-4,4]$ the AIC is not favoring any of the models (or the preference is slight), and if it lies outside of $[-7,7]$ then the model is strongly favored (opposed) by the AIC.
Notice that in the RNSI model, we have three extra parameters, because in the $\Lambda$CDM model we fixed the standard neutrino parameters to $\sum m_{\nu}=0.06$ eV and $N_{\rm eff}=3.046$.
In table \ref{tab:limits}, we show the $\Delta \rm{AIC}$ values for the different masses and datasets.
We observe when using Planck + BAO, $\Lambda$CDM is slightly preferred.
However, when $H_0$ is included, RNSI is favored over the cosmological standard model, particularly, for the cases $m_\varphi=0.1$ eV and $m_\varphi=1$ eV.

The results discussed in this section show that the RNSI model offers interesting features that are worth continuing to study.
This model emerges as a good candidate to ease the Hubble tension, with a strong overall fit, while also requiring physics beyond the standard model of particles.

\begin{table*}
  \centerline{
\begin{tabular} {| l | c c | c c| c c|}
\hline
Data: Planck + & BAO & BAO+H0  & BAO & BAO+H0 &  BAO & BAO+H0 \\
\hline
    $m_{\varphi}[\rm{eV}]$  &\multicolumn{2}{c|}{0.5}  &\multicolumn{2}{c|}{5} &\multicolumn{2}{c|}{50}\\
    \hline
$H_0 [\rm{km\ s^{-1}/Mpc}]     $ & $68.0\pm 1.4  $ & $71.29\pm 0.84        $ & $67.8^{+1.0}_{-1.3}   $ & $71.03^{+0.82}_{-0.91}$ & $67.5\pm 1.3 $ & $70.93\pm 0.83$ \\
$N_{\rm eff}              $ & $3.08\pm 0.23 $ & $3.59^{+0.16}_{-0.15} $ & $3.05^{+0.18}_{-0.22} $ & $3.55\pm 0.15         $ & $2.98\pm 0.21$ & $3.52\pm 0.15 $ \\
$\sum m_{\nu}[\rm{eV}]         $ & $< 0.126      $ & $< 0.112              $ & $< 0.104              $ & $< 0.112              $ & $< 0.107     $ & $< 0.123      $ \\
$g_{\nu}\times 10^{14}    $ & $< 22.9       $ & $18.7^{+5.4}_{-4.8}   $ & $< 15.6               $ & $13.4^{+6.4}_{-5.8}   $ & $< 65.8      $ & $< 67.9       $ \\
$\Delta$AIC  & 8.04 &-9.18 & 5.52 & -7.64 & 8.9 & -2.94 \\
\hline
\end{tabular}}
\caption{\label{tab:limits2} Observational limits for different models with varying $N_{\rm eff}$ and $\sum m_\nu$. The upper limits are expressed at $2\sigma$ while the rest of the limits are at $1\sigma$.
We obtain a positive value of the interaction coupling for $m_\varphi$ corresponding to 0.5 and 5 eV.}
\end{table*}

\section{\label{sec:conclusions} Conclusions and perspectives}

Along this paper, we have explored the cosmological implications of the existence of scalar-mediated resonant neutrino self-interactions contributing to a neutrino-neutrino elastic scattering. This corresponds to a previously unexplored range of mediator masses, where the
resonance enhances substantially the cross section. The cosmological observables become sensitive to very small couplings, especially when the associated peak occurs close to recombination. This provides strong bounds for interactions with scalar masses in the regime between sub-eV to hundreds of eV. 

For our study, we employed a modified version of \textsc{CLASS} \cite{class2,class4} and \textsc{MontePython} \cite{Montepython2013,Brinckmann2019_montepython} with public data of the CMB by Planck 2018, BAO and local measurements of $H_0$.
We explored a sample of fixed mediator masses in the range $10^{-2}$ eV $\leq m_{\varphi}\leq 10^{2}$ eV, where we varied the interacting coupling $g_{\nu}$,
the neutrino parameters $N_{\rm eff}$ and $\sum m_{\nu}$, and the base $\Lambda$CDM parameters.

When using the Planck + BAO data batch, the tension with local measurements of $H_0$ is reduced significantly from 4.9$\sigma$ in the standard cosmological model down to values between 3.3$\sigma$ and 2.8$\sigma$.
Nevertheless, to reach those values the model requires extra radiation, which by itself requires a RNSI consistent model with extra radiation and a mechanism to avoid other constraints (see for example \cite{Huang2021}).
Consequently, we argue the that resonant NSI model could be a good candidate to solve the $H_0$ tension.
Furthermore, when adding local $H_0$ data, the model accepts large $H_0$ values with extra radiation and nonnull interaction couplings.
Actually models with masses in the range $0.5$ eV $\leq m_{\varphi}\leq 10$ eV prefer nonzero interactions with 2.3$\sigma$ to 3.9$\sigma$ levels of significance.
However this last result was observed combining data that are in tension at high statistical significance yet.
The search for such small couplings in neutrino experiments could take decades to plan, even so, this region of the parameter space is reachable using cosmological (or astrophysical) data.
We did not observe significant deviations or correlations on other cosmological parameters.

The AIC favored our model over $\Lambda$CDM when we used Planck + BAO + $H_0$ data. This means that the model improved the likelihood enough to justify the additional parameters. When we used only Planck + BAO data the criterion prefers the $\Lambda$CDM model, which indicates that the preference in the full dataset comes from the ability of our model to reduce the $H_0$ tension.

We argue that future projects can improve our current analysis in several ways.
First, we ignored the neutrino mass in the computation of the interaction rate, a more precise computation should include it. 
This will add uncertainty to the analysis, especially for the cases where the mediator mass is below the atmospheric neutrino mass scale $\sim 0.05$ eV.
Second, we ignored the resonant production of scalars and the neutrino decay (inverse decays) into lighter neutrinos and scalars.
These processes would be important because they may change the neutrino background, perturbation evolution, and even compromise the validity of the relaxation time approximation \cite{Oldengott_2015,Oldengott_2017}.
Future studies in RNSIs should pursue a joint analysis that includes both inelastic and elastic processes.
From the data analysis point of view, it would be interesting to test the model with new datasets.
For this matter, future cosmological surveys will help to untangle the parameter constraints (see for instance \cite{blum2022snowmass2021,rhodes:hal-02915474}).
Additionally, we should track the model comparison using Bayesian methods to discriminate against other cosmological models.

The physics of neutrino nonstandard interactions may change our understanding of the Universe.
Exploring all the available parameter space could open new windows to solve current cosmological problems such as the $H_0$ tension.
The previously unexplored region studied in this work now brings interesting links to physics beyond the standard model of particle physics. In particular, the preference in certain masses for a nonzero interaction is intriguing and should be studied further.

\begin{acknowledgments}
The authors thankfully acknowledge Cl\'uster de Superc\'omputo Xiuhcoatl (Cinvestav) for the allocation of computer resources.
We thank Mauricio L\'opez for the permission to use his Markov chains for the noninteracting case.
We thank an anonymous referee for the insightful comments that helped to improve our discussion.
Work partially supported by Conahcyt, M\'exico, under FORDECYT-PRONACES Grant No. 490769. 
G.G.A. is grateful to FORDECYT PRONACES-CONAHCYT Grant No. CF-MG-2558591, FOSEC SEP-CONAHCYT Ciencia B\'asica A1-S-21925, FORDECYT-PRONACES-
CONAHCYT 304001 and UNAM-DGAPA-PAPIIT IN117723.
J.V. acknowledges Conahcyt, for partial support, under the project within ``Paradigmas y Controversias de la Ciencia 2022'', No. 319395.
\end{acknowledgments}

\appendix
\section{Further details on cosmological constraints}\label{appendix}

\begin{figure}
    \centering
    \includegraphics[width=.42\textwidth]{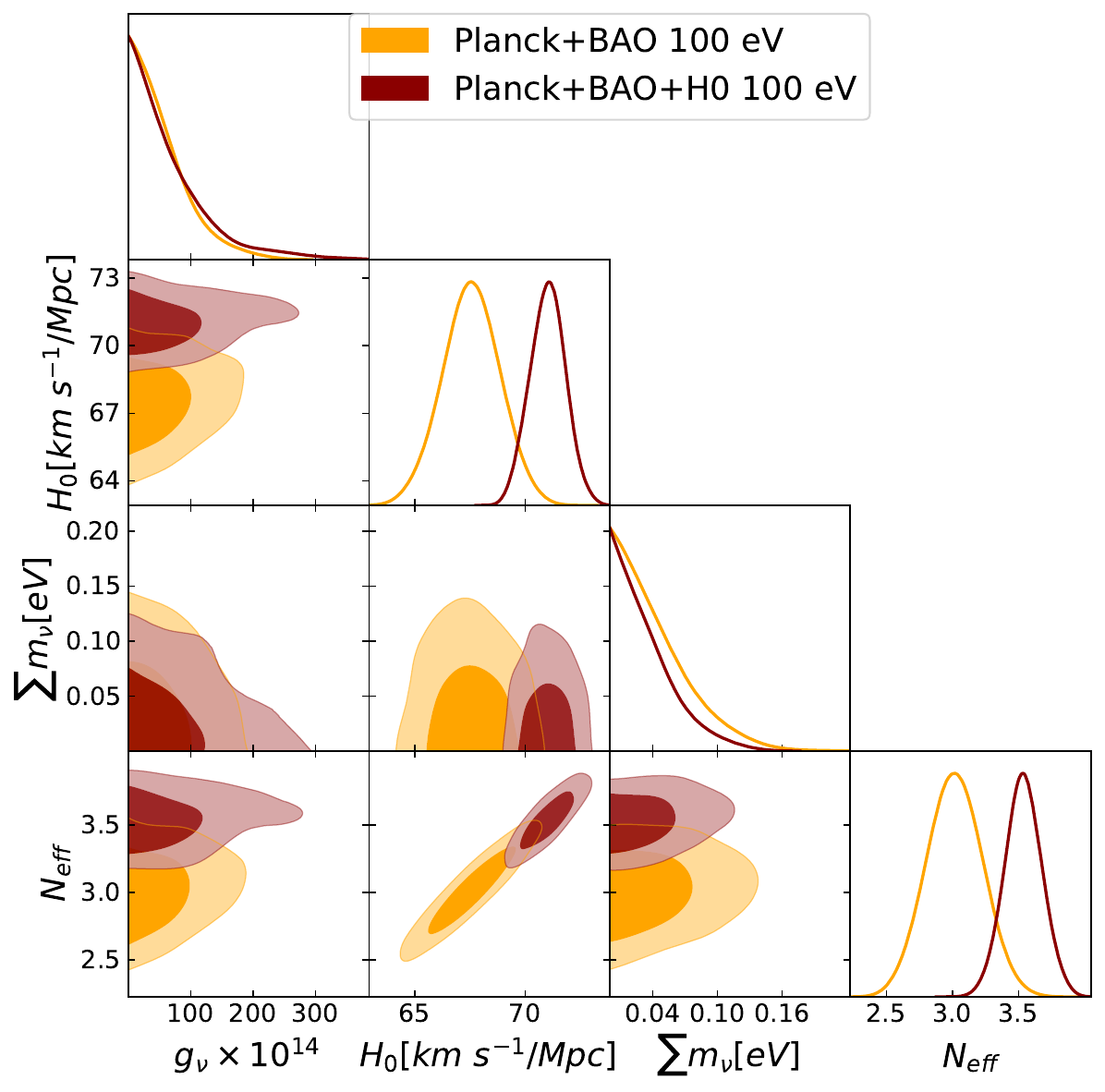}
    \caption{ The plot displays the 68\% and 95\% confidence regions for the RNSI model with a fixed mediator mass of $m_\varphi=100$ eV, constraints are for the Planck+BAO and Planck+BAO+$H_0$ datasets.}
    \label{fig:triangular100ev}
\end{figure}

\begin{figure}
    \centering
    \includegraphics[width=.42\textwidth]{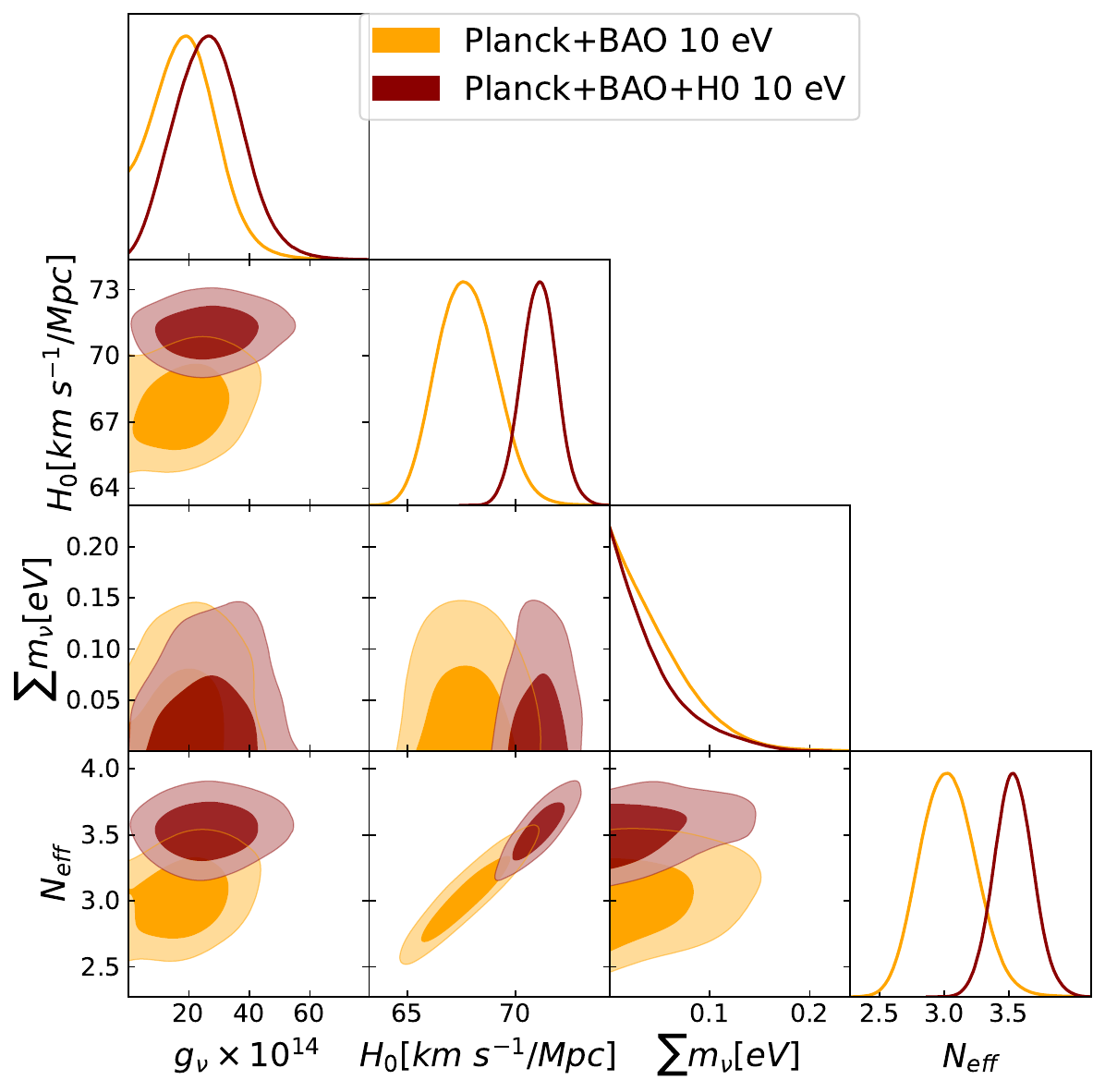}
    \caption{Posterior probability for different sets of data and for $m_\varphi=10$ eV.}
    \label{fig:triangular10ev}
\end{figure}

\begin{figure}
    \centering
    \includegraphics[width=.42\textwidth]{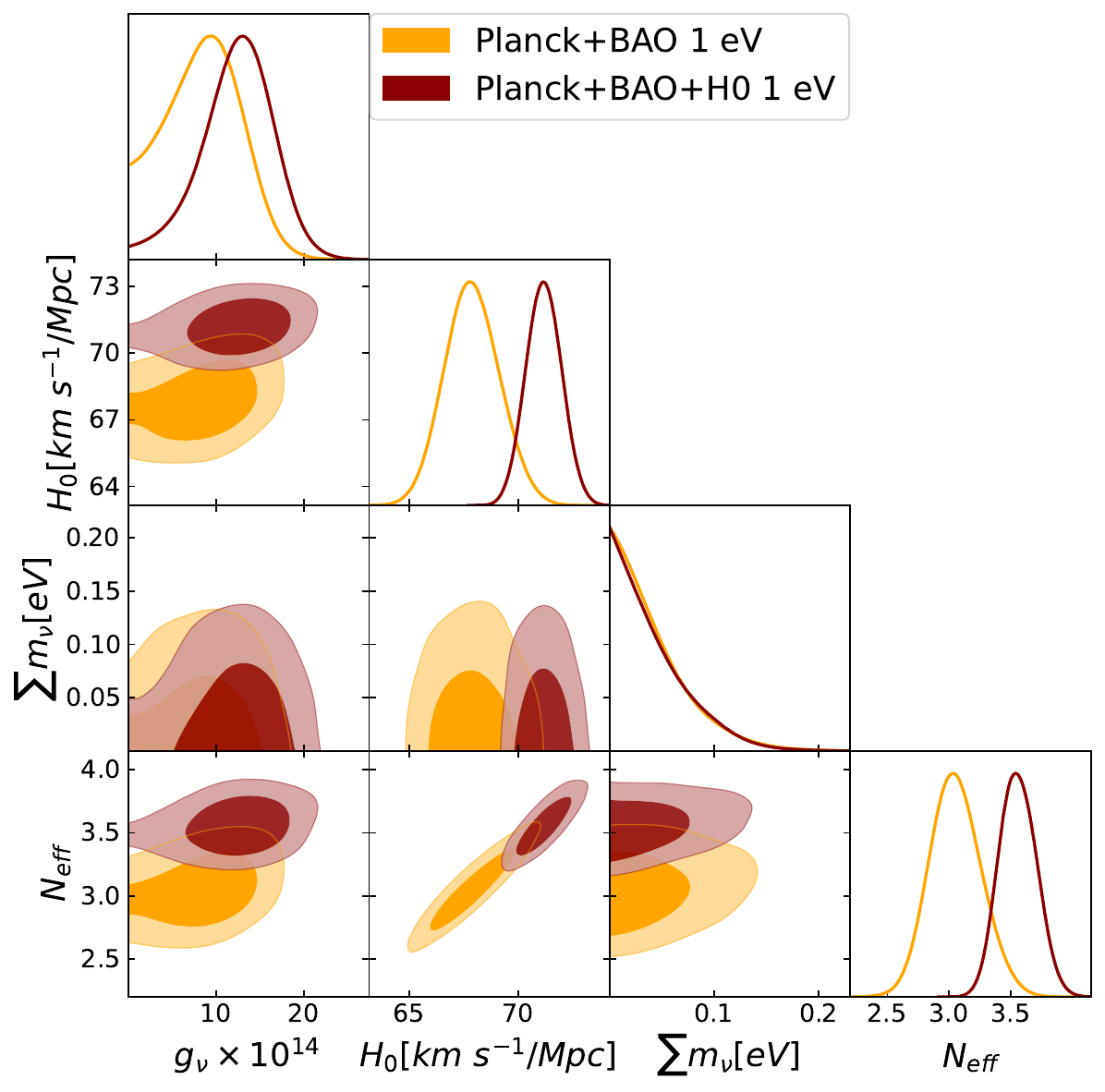}
    \caption{Posterior probability for different sets of data and for $m_\varphi=1$ eV.}
    \label{fig:triangular1ev}
\end{figure}

\begin{figure}
    \centering
    \includegraphics[width=.42\textwidth]{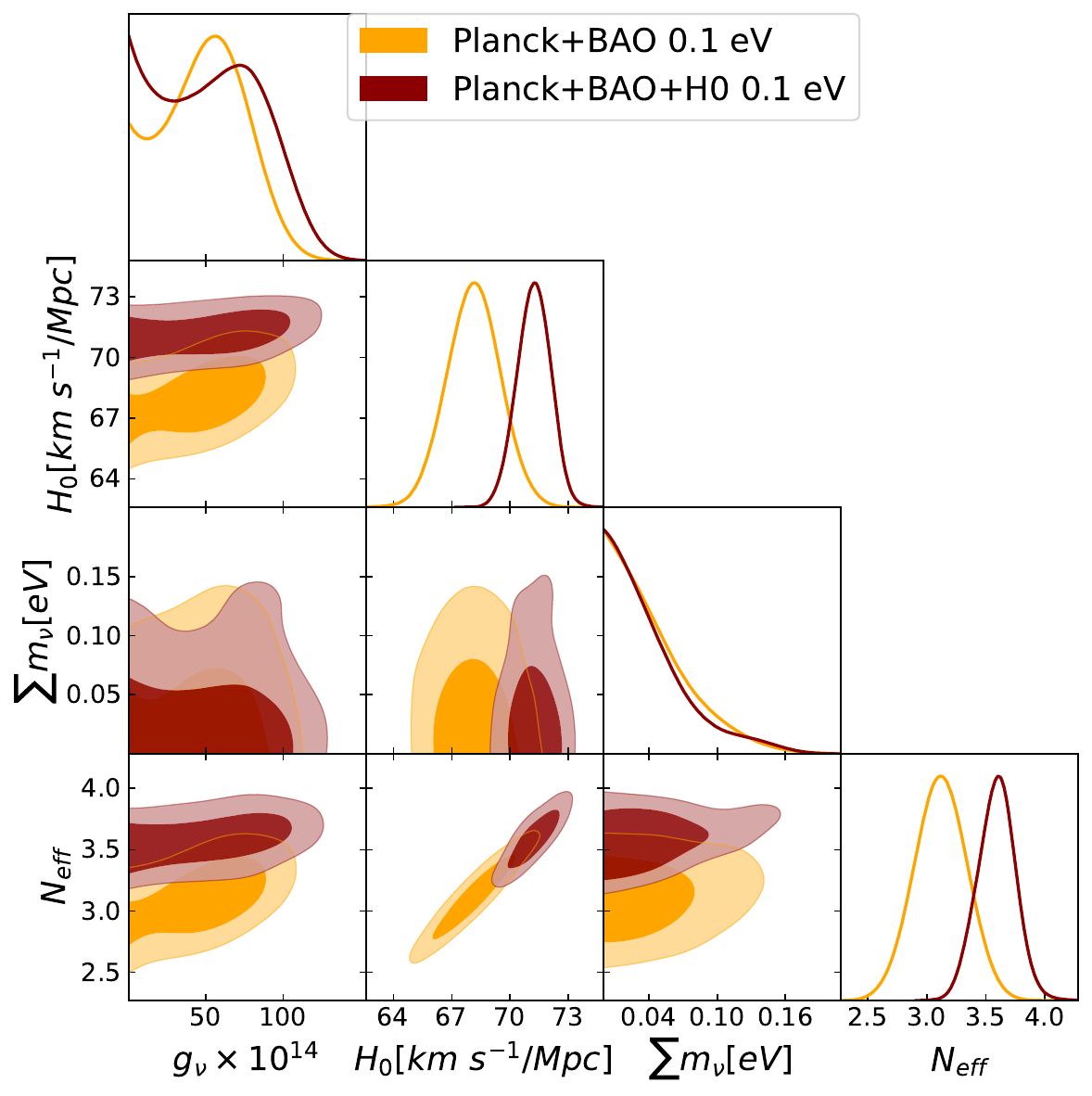}
    \caption{Posterior probability for different sets of data and for $m_\varphi=10^{-1}$ eV.}
    \label{fig:triangular10minus1ev}
\end{figure}
\begin{figure}
    \centering
    \includegraphics[width=.42\textwidth]{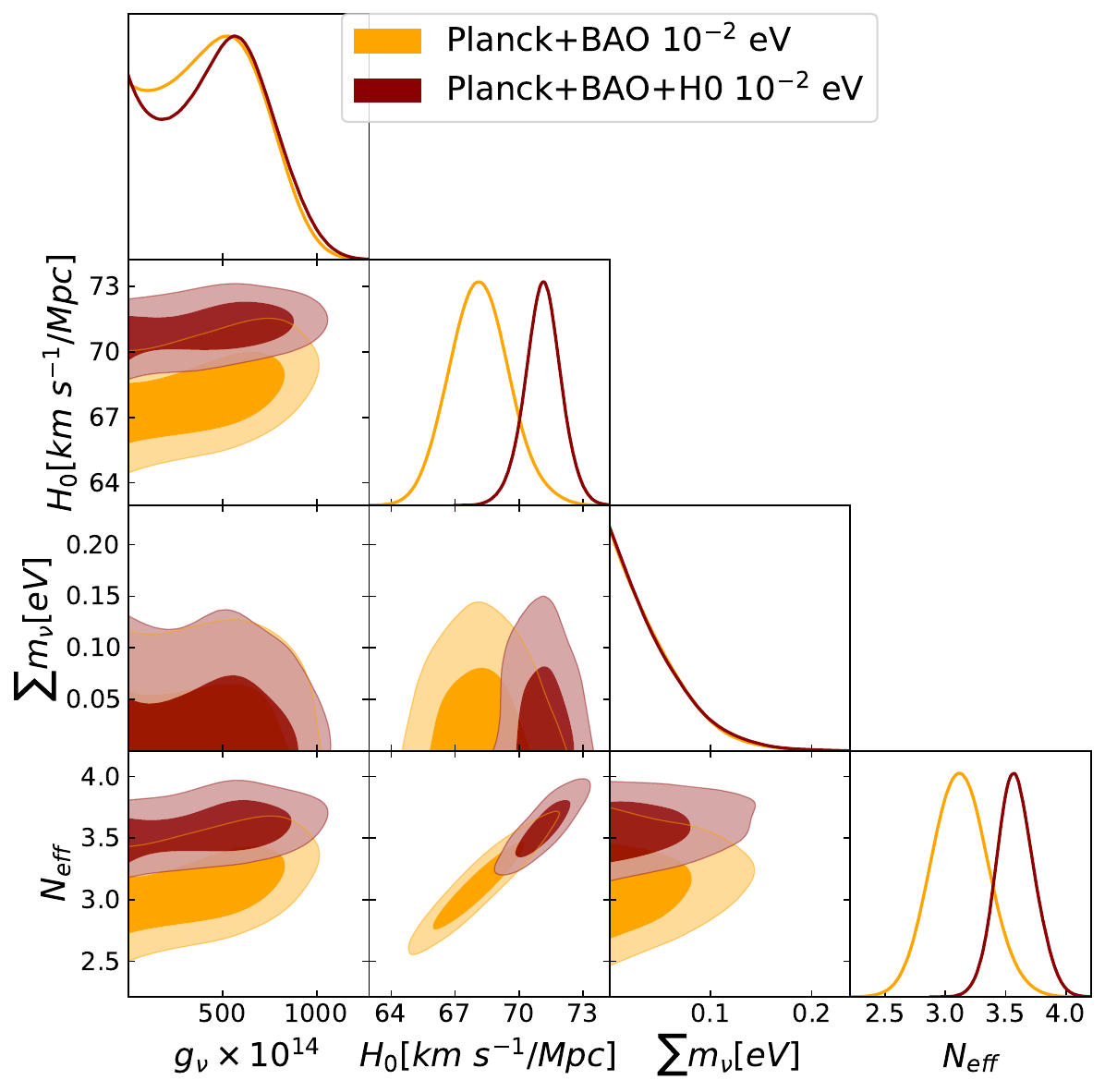}
    \caption{Posterior probability for different sets of data and for $m_\varphi=10^{-2}$ eV.}
    \label{fig:triangular10minus2ev}
\end{figure}

This appendix provides a discussion of some details about the cosmological results.
First, we focus on the constraints of the base parameters.
We varied the usual six $\Lambda$CDM  parameters $\Omega_b$, $\Omega_{cdm}$, $\theta_s$, $A_s$, $n_s$ and $\tau_{reio}$, and found that the results are consistent with the cosmological standard model constraints across all mediator masses.
As an example, when using $m_{\varphi}=1$ eV for Planck + BAO, we obtained: 
 $\Omega_bh^2=(2.2428\pm0.0188)\times10^{-2}$, $\Omega_{\rm cdm}h^2=0.1194\pm0.0034$, $100\theta_{s}=1.0420\pm0.0006$, $\ln10^{10}A_{s}=3.0466\pm0.0168$, $n_s=0.9690\pm0.0081$, and $\tau_{\rm reio}=(5.7724\pm0.7546)\times10^2$. 
On the other hand, when we include $H_0$, the constraints (also for $m_{\varphi}=1$ eV) are $\Omega_bh^2=(2.2808\pm0.0149)\times10^{-2}$, $\Omega_{\rm cdm}h^2=0.1266\pm0.0028$, $100\theta_{s}=1.0410\pm0.0005$, $\ln10^{10}A_{s}=3.0675\pm0.0167$, $n_s=0.9865\pm0.0062$, and $\tau_{\rm reio}=(6.1376\pm0.8211)\times10^2$.  
Similar bounds were obtained for the other mediator masses.

Now we focus on the degeneration of the RNSI parameters.
In Figs. \ref{fig:triangular100ev}--\ref{fig:triangular10minus2ev}, we show the 2-D posteriors for the neutrino parameters and $H_0$.
In all cases, we observe similar correlations.
For instance, there is a positive correlation between $g_\nu$, $N_{\rm eff}$, and $H_0$.
The strongest correlation is between $N_{\rm eff}$ and $H_0$, this occurs also in the standard neutrino scenario. 
The correlation of $g_\nu$ with these parameters is moderate, however, it is steeper in the Planck + BAO dataset.
Thus, from these we confirm the expected RNSI effect, which prevents neutrinos from free streaming, which the model compensates with a larger $N_{\rm eff}$, and due to correlations, resulting in a larger $H_0$ value, this is sharpest when $H_0$ forms part of the data.

The correlation of $\sum m_\nu$ with the other parameters is mild.
Except, with $N_{\rm eff}$, which presents a positive and moderate correlation.
We conclude that the neutrino mass has a very modest impact on the final constraints of the other parameters and therefore its bounds are similar to those obtained in the standard scenario.

Now, we discuss the results of the three extra mediator masses.
In table \ref{tab:limits2}, we show the constraints to the parameters for the extra values of $m_\varphi$.
Within these masses ($0.5,5$ and $50$ eV) we observe an interpolation of the results when using the main mass set (also see Fig. \ref{fig:gnu_vs_mass}).
Specifically, we run these extra MCMCs to observe the behavior of the $g_\nu$ bound with all the datasets.
We can see that $0.5$ and $5$ eV are consistent with a nonzero value, and, for $50$ eV, we do not observe any preference for nonstandard physics.

Regarding the extra $\Delta \rm{AIC}$ values
(Eq. \eqref{eq:aic}), our model is penalized when considering Planck + BAO.
Although, for the whole batch, our model is favored by this criteria.
These extra points strengthen our confidence in the results we report on the main text.
Furthermore, concerning the results of $H_0$, we obtain similar results to the main sample, in which the tension is better reduced for a lighter mediator.

\bibliography{references}

\end{document}